\documentclass{emulateapj}
\newcommand{\bm}[1]{\ensuremath{\mbox{\boldmath $#1$}}}
\newcommand{\bmm}[1]{{\mathbf{#1}}}

\newcommand{\rmn}[1] {{\rm #1}}

\newcommand{\text}[1] {{\textrm{#1}}}
\newcommand{\eqref}[1] {equation $($\ref{#1}$)$}
\def\beq{\begin{equation}}
\def\eeq{\end{equation}}
\def\beqn{\begin{eqnarray}}
\def\eeqn{\end{eqnarray}}

\def\h{\mathrm{h}}
\def\d{\rmn{d}}
\def\pa{\partial}
\def\ba{\bm{\alpha}}
\def\fracj#1#2{{\textstyle{#1\over#2}}}
\def\bxi{\bm{\xi}}
\def\ti{\widetilde}
\def\O{\Omega}
\def\Om{\ensuremath{\Omega_{\mathrm{m}}}}
\def\o{\omega}

\def\2gcm{\textrm{g cm$^{-2}$}}
\def\Scr{\Sigma_{\mathrm{crit}}}

\def\av#1{\l \langle{#1}\r \rangle}
\def\l{\left}
\def\hmpc{\:\mathrm{h}^{-1}\mathrm{Mpc}}
\def\r{\right}
\def\sg{\sigma}
\def\Sig{\Sigma}
\def\cf{{\cal F}}
\def\k{\kappa}
\def\P{{P}}
\def\pnl{{P}_{{\!\textrm{\tiny NL}}}}
\def\dnl{\Delta_{\text{\scriptsize NL}}}
\def\kmin{\k_{\mathrm{min}}}
\def\kmax{\k_{\mathrm{max}}}
\def\ktot{\k_{\mathrm{tot}}}

\def\arcsec{$^{\prime\prime}$}
\shortauthors{Sudeep Das and Jeremiah P. Ostriker}
\shorttitle{ Gravitational Lensing Probabilities}
\begin{document}
\title{Testing a new  analytic model for gravitational lensing probabilities}
\author{Sudeep Das and Jeremiah P. Ostriker}
\email{sudeep, jpo@astro.princeton.edu}
\affil{Princeton University Observatory, Princeton, NJ 08544-1001}
\begin{abstract}
We study gravitational lensing with a multiple lens plane approach, proposing a simple analytical model for the probability distribution function (PDF) of the dark matter convergence, $\kappa$, for the different lens planes in a given cosmology as a function of redshift and smoothing angle, $\theta$. The model is fixed solely by the variance of $\kappa$, which in turn is fixed by the amplitude of the power spectrum, $\sigma_8$.  We test the PDF against a high resolution Tree-Particle-Mesh  simulation and find that it is far superior to the Gaussian or the lognormal, especially for small values of $\theta \ll 1'$ and at large values of $\kappa$ relevant to strong lensing. With this model, we predict the probabilities of strong lensing by a single plane or by multiple planes. We find that for $\theta \sim 10''$, a single plane accounts for almost all ($\sim 98\%$) of the strong lensing cases for source redshift unity. However, for a more typical source redshift of $4$, about $12\%$ of the strong lensing cases will result from the contribution of a secondary clump of matter along the line of sight, introducing a systematic error in the determination of the surface density of clusters, typically overestimating it by about $2-5\%$.  We also find that matter inhomogenieties introduce a dispersion in the value of the angular diameter distance about its cosmological mean. The probable error relative to the mean increases with redshift to a value of about $8\%$ for $z\simeq 6$ and $\theta \sim 10''$. 
\end{abstract}
\section{Introduction}{\rm }

Gravitational lensing provides a unique tool for studying the cosmological distribution of matter, because it directly probes the gravitational potential and is indifferent to the nature or the physical state of matter. It is, therefore, free from the assumptions that can plague other dynamical methods for studying gravitational fields.\par
Tidal deflections of light rays by structures along its path can weakly distort and magnify distant sources. These effects, although small on the individual level, can be statistically significant for an ensemble of sources. Thus, statistical properties of the intervening matter distribution can be inferred from such {\it weak lensing} studies {\citep[see, for e.g.][for reviews]{ blandford_narayan_1992, bartelmann_schneider_2001, refregier_2003}}. This field is extremely promising at the present time but is fraught with observational challenges, because these subtle effects have to be very delicately isolated from comparable systematic errors.\par 
Occasionally, the distortion due to the intervening matter may be strong enough to produce multiple images of the source. {\it Strong lensing} is extremely sensitive to the cosmology and hence, can be used to constrain cosmological parameters. Small changes in the cosmological parameters can significantly alter the nonlinear structure formation history of the universe and thereby result in huge differences in the abundance of compact objects which behave as potential strong lenses. This makes the probability of any specific type of strong lensing event a useful observable for constraining cosmology {\citep[see, for example,][]{TOG,narayan_blandford_1991,kaiser_1992,CGOT,nakamura_suto_1997,cohn_etal_2001,keeton_2001,premadi_2001,oguri_etal_2002,li_ostriker_2002,li_ostriker_2003}}. For example, the statistics of lensed quasars or galaxies lensed by clusters into giant arcs can be used to study the population of lenses, their distribution in redshift space, mass, and hence the cosmological model. Strong lensing systems may also be studied individually, in order to obtain a detailed understanding of the structure and the distribution of matter in the lensing object and the source {\citep[see, for example,][]{claeskens_etal_2000,broadhurst_etal_2000,shapiro_iliev_2000, cohn_etal_2001, belokurov_etal_2001, patnaik_narasimha_2001,inada_etal_2003, fassnacht_etal_2005, sluse_etal_2005, wayth_etal_2005, claeskens_etal_2006}}. In this paper, we shall be mostly interested in the former problem, namely the statistics of strong lensing events.\par
In the absence of satisfactory analytic tools, the most widely used method for studying strong lensing statistics has been the analysis of large numerical simulations. The method of tracing the path of light rays through a simulated universe evolving in accordance with some specific cosmological model is being extensively used to study the multitude of lensing phenomena. For example, giant arc statistics using ray tracing simulations of clusters in a $\Lambda$CDM  cosmology are being thoroughly investigated \citep[examples include][etc.]{ bartelmann_etal_1998, WBO_2004a,dalal_holder_hennawi_2004,ho_white_2004,li_etal_2005,hennawi_etal_2005a,hennawi_etal_2005b}.  However, these simulations are costly in that they require extensive computing power and time. This also restricts the range of cosmological models that can be investigated; in most cases only the currently favored cosmological model is used. One must not, however overlook the array of important analytical works from the so-called  halo model approach \citep[see, for e.g.][]{li_ostriker_2003,oguri_etal_2003, oguri_keeton_2004}, which allow accurate computation of lensing probabilities for a single lens.  \par
The goal of this paper is to provide an analytic method to address some of the questions pertaining to gravitational lensing statistics that, at the cost of having much more limited information available than from detailed numerical studies, would have the benefit of allowing any definite cosmological model to be investigated. We test our results against numerical simulations performed by the Tree-Particle-Mesh (TPM) code \citep{bode_ostriker_2003} for a $\Lambda$CDM universe. The method lends itself easily to the treatment of lensing under other cosmologies.\par
We consider the fluctuations in the matter density along a light cone of a given  opening angle.  We propose an analytic method to compute the probability that  a light beam will encounter a given surface density while passing through a slice of matter centered around a given redshift. The model is completely defined by the variance of the projected density of the matter in the slice, smoothed on the angular scale corresponding to the beam. We show how the variance of surface mass density on the lens can be predicted from the power spectrum in Section~\ref{sec:variances} and develop the analytic model in Section~\ref{sec:model}.

Armed with this result, we adopt a multiple lens plane approach to study strong lensing. We divide up the line of sight distance to the source into contiguous slices of equal comoving depth, each of which is chosen to be larger than the correlation length so that each slice can be considered statistically independent of the others. We imagine the matter inside each of these slices to be projected onto a plane at its center. The probability density function (PDF) for the surface density on each plane is then constructed with our analytical method. The matter encountered by the light ray while propagating through the universe is then simulated by randomly picking a surface mass density from the PDF on each slice it passes through. In fact, the surface density scaled by the so called critical density required for multiple image formation, better known as the convergence, is computed on each of these planes. We identify different strong lensing scenarios according as whether the individual or the additive value of convergences so obtained is in excess of unity. We shoot a statistically significant number of rays to obtain the probabilities of different events. In particular, we study the probability of strong lensing as a function of source redshift and angular scale.\par
Another important question is the role of auxiliary planes aiding the main lens to produce multiple images. It is known that some multiply imaged quasars \citep[e.g. the quadrupole system B1422+231, see][]{kormann_etal_1994} cannot be well modeled under the thin lens approximation, which assumes all mass responsible for lensing to be concentrated on a mathematical surface of zero thickness \citep{SEF_1992}. Also, observations suggest cases where galaxies at disparate redshifts appear to have contributed to the lensing event \citep{tonry_1998, augusto_etal_2001, chae_etal_2001,johnston_etal_2003,fassnacht_etal_2005}. In order to study the probabilities of such events, we identified, within the strong lensing cases, the ones that were caused by a single plane or by the concordance of two or more planes and studied their probabilities separately. Incidentally, a similar study was done in \citet{WBO_2004b}, who studied the phenomenon at the pixel level ($\sim 1.5''$) and reported a significant contribution of auxiliary planes in the strong lensing probabilities. With our method, we can conveniently include an element of angular smoothing depending on the strong lensing situation and study the probabilities as a function of smoothing radius. We treat this problem in Section~\ref{sec:prob}. \par
 We also note that our model works very well even in the weak lensing regime (angles $\ga 1'$). The statistics of weak lensing convergence can be studied from the  magnification effects of clustered matter which leads to variations in the number density and image sizes of galaxies across the sky \citep[see, e.g.,][]{jain_etal_2002}. Also, shear maps generated from weak lensing surveys yield statistical properties of the convergence. Thus a good analytic model for the convergence can be useful in constraining the properties of large scale structure \citep{barber_etal_2004}, although the practicalities involved may seem rather daunting at the moment \citep{bartelmann_schneider_2001}. The situation will improve with forthcoming surveys having high signal-to-noise ratios, enabling us to study the small scale clustering of matter.\par
In Section~\ref{sec:mass}, we study the effect of the chance accumulation of secondary matter along the line-of-sight on the estimation of cluster masses and finally, in Section~\ref{sec:angdia}, we examine the dispersion in the angular diameter distances due to inhomogeneous distribution of matter.
\section{Variance in projected overdensities}
\label{sec:variances}
  Since the fluctuations in matter density along the line of sight parallel the growth of structure in  the universe, with higher redshifts corresponding to an earlier era in the structure formation history, these variances are intricately related, and can, in fact, be predicted from the evolution of the matter power-spectrum. For large scales or early times the growth of structure is linear and the power-spectrum is linearly related to its primordial form  through a transfer function \citep[see][]{bbks_1986}. However, on small scales and late times, the growth of structure becomes nonlinear and small scale modes couple and grow more rapidly than in the linear regime. The evolution of the power-spectrum in this era cannot be calculated analytically with sufficient accuracy and one has to resort to numerical simulations to find some analytic fitting function for the simulated power spectrum. This was first done by \cite{peacock_dodds_1996} who used N-body simulations  based on hierarchical clustering of matter to produce an analytic fit. Later,  \cite{smith_etal_2003} proposed a halo model \citep{seljak_2000, ma_fry_2000} inspired fitting function which was shown to be in excellent agreement with numerical simulations \citep{barber_taylor_2003} in the context of the lensing convergence power-spectrum. We have adopted the \citet{smith_etal_2003} fitting functions in this work. In order to calculate the variances in the 2D matter distribution obtained by projecting the matter along the line of sight in a redshift slice, the 3D power-spectrum has to be related to its 2D counterpart. This is done by the well known Limber approximation \citep{kaiser_1998}. Using this method, we show that the 2D  variances and consequently, the probability distribution for surface density can be predicted with sufficient accuracy given a cosmological model.\par
In the present approach, we divide the matter along the line of sight from the observing   redshift of zero to a redshift of $6.34$ into $38$ slices, each $160\hmpc$ thick, and project the matter in each of these slices onto a plane at its center. Therefore we have lens planes at $80+j\times 160\: \hmpc$ for $j=0,..,37$. Let  $\Sig(\bm{\xi},z)$ denote the projected surface mass density at a position $\bm{\xi}=\{\xi_1,\xi_2\}$ on a plane at redshift $z$. We denote the same quantity, when smoothed by a Gaussian window of angular radius $\theta_0$, by $\Sig_{\theta_0}(\bm{\xi},z)$. We are interested in finding the variance $\sigma_2^2(\theta_0,z)\equiv \av{\delta_2^2(\bm{\xi};\theta_0)}$ in the projected overdensities $\delta_2(\bm{\xi})$ given by, 
\beq
\delta_2(\bm{\xi};\theta_0)=\frac{\Sigma_{\theta_0}(\bm{\xi},z)-\av{\Sigma_{\theta_0}}}{\av{\Sigma_{\theta_0}}}.
\eeq
Here and henceforth, the subscript $2$ on any symbol is meant to assert that it is a quantity depending on  2-dimensional variables. 
In order to obtain $\sigma_2^2(\theta_0,z)$, we first use the Limber Equation \citep{kaiser_1998} to obtain the 2-D power spectrum for the distribution of $\Sigma$ on a plane from the nonlinear 3D power spectrum $\pnl(\bmm{k})$. Let us denote the comoving distance to the center of the bounding planes of the $i$th slice by $\chi_{i-1}$ and $\chi_i$, so that the first slice is bounded planes at by $\chi_0$ and $\chi_1$ and so on. The coordinate of a point inside the $i$th slice can be denoted by, 
\beqn
\nonumber
\bm{x}(\chi,\bm\theta)&=\chi(z)(\theta_1,\theta_2,1),&\;\;\;\; (\chi_{i-1}< \chi <\chi_i),\eeqn
 where $\theta_1$ and $\theta_2$ are two dimensionless numbers (angles) specifying the position of the point on the plane orthogonal to the line of sight. Since we are interested only in the projection of the matter inside this slice, the overdensity seen at the point $(\theta_1,\theta_2)$ on the plane of projection is,
\beq 
\delta_2(\bm{\theta})=\int \d\chi W(\chi) \delta[\bm{x}(\chi,\bm\theta)],
 \eeq
where $W$ is a selection function, which, in our case, is given by a step function,
\beq
W(\chi)=
\left\{{\begin{array}{cr}
\frac{1}{(\chi_i-\chi_{i-1})} & \text{if   } (\chi_{i-1}<\chi<\chi_i)\\
 0 & \text{otherwise.}
\end{array}}\r.
\eeq
The Fourier space conjugate to $\bm{\theta}$ is denoted by ${\bm{l}}$, and defined through,
\beq
\delta_2(\bm{l})=\int \d^2\bm{\theta} e^{i{\bm{\scriptstyle{l}}\cdot \bm{\scriptstyle{\theta}}}}\delta_2(\bm\theta).
\eeq 
 The 2D power-spectrum is  given by the standard Limber approximation \citep{kaiser_1998,dodelson_2003}, 
\beqn
\P_2{( l)}&=&\int \d\chi \frac{W^2(\chi)}{\chi^2}\pnl(l/\chi,z)\\
&=& \frac{1}{(\chi_i-\chi_{i-1})^2}\int_{\chi=\chi_{i-1}}^{\chi_i} \d\chi \frac{ \pnl(l/\chi)}{\chi^2},
\eeqn 
using our selection function.\par
By introducing $k\equiv l/\chi$, we can rewrite this as,
\beq
\P_2{(l,z)}=\frac{1}{l(\chi_i-\chi_{i-1})^2}\int_{ l/\chi_i}^{l/\chi_{i-1}}\d k  \pnl(k,z).
\eeq
Now, the variance of the projected overdensity, Gaussian smoothed at an angle $\theta_0$ can be calculated by using, 
\beqn
\sigma_2^2(\theta_0,z_i)&=&\frac{1}{(2\pi)^2}\int \Lambda^2(l\theta_0) \P_2({l})\d^2\bm{l},
\eeqn
where $\Lambda(l,\theta_0)$ is the Fourier transform of the ${\bm{\theta}}$ space Gaussian smoothing window of width $\theta_0$, and is given by, 
$$\Lambda(l,\theta_0)= e^{- l^2 \theta_0^2/2}.$$
With this we have, 
\beqn  
\sigma_2^2(\theta_0,z_i)&=&\frac{1}{(2\pi)^2} \int \d^2\bm l \: e^{-{l^2 \theta_0^2}} \P_2(l)\\
\nonumber &=&\frac{1}{2\pi(\chi_i-\chi_{i-1})^2}   \int \d l\: e^{-{l^2 \theta_0^2}}  \int_{ l/\chi_i}^{l/\chi_{i-1}}\d k  \pnl(k,z_i)\\
\label{final}
&=&\frac{\pi}{(\chi_i-\chi_{i-1})^2}   \int \d l\: e^{- {l^2 \theta_0^2}}  \int_{ l/\chi_i}^{l/\chi_{i-1}}\frac{\d k}{k^3}  \dnl^2,
\eeqn
where $\dnl^2=\frac{1}{2\pi^2}\pnl(k) k^3 $ is the dimensionless power spectrum. \par
The integral above requires the value the non-linear power spectrum $\dnl^2(k)$ as a function of scale $k$ and redshift $z$. Incidentally, this can be generated using the routine ``{halofit}'' by Smith and Peacock \citep[see][]{smith_etal_2003}. 
\par We test   our predictions against a high resolution cosmological simulation using the Tree-Particle-Mesh (TPM) code \citep{bode_ostriker_2003}.  The simulations were performed in a box with a comoving side length of  $L=320 h^{-1}$Mpc. We used $N=1024^3=1,073,741,824$ particles, so the individual particle mass is $m_{\rm p}=2.54\times 10^9 h^{-1}$ M$_\odot$. The cubic spline softening length was set to $\epsilon=3.2 h^{-1}$ kpc, producing a ratio of    box size to softening length of $L/\epsilon =10^5$. The output was stored at 19 redshift values out to $z \approx 6.4$, such that the centers of the saved
boxes matched comoving distances of $(160 + n \times 320) \hmpc$, where $n=0,...,18$. Each box was then divided into $9\times 9$ separate square cylinders along its length.  This was done for three orthogonal projections, leading to 243 sub-volumes. Two lens planes were produced for each sub-volume by bisecting along the line-of-sight and projecting the mass in each $160 h^{-1}$Mpc--long volume onto a plane. {Thus, we have planes at comoving distances of $(80+j\times 160) \hmpc$, where $j=0,...,37$.} At the highest redshift, each plane has a side length of
$35.6 h^{-1}$Mpc and contains $800^2$ pixels, making the pixel size $44.4 h^{-1}$kpc comoving.  At lower redshifts, an opening angle of about $20''$ was maintained by keeping  the number of pixels constant but progressively decreasing the size of the planes.  Thus in the lowest redshift box the lensing planes are $1.9 h^{-1}$Mpc on a side and the pixel size is $2.3 h^{-1}$kpc, which is slightly lower than our spatial resolution.  Because at lower redshifts the plane does not cover all of the corresponding sub-volume, there is a random offset
perpendicular to the line of sight within each sub-volume. In all lensing applications discussed below, we will not consider structures smaller in size than 5 pixels or three times our spatial resolution. \par
In order to test the results against the WMAP normalized \citep{spergel_etal_2003} $\Lambda$CDM TPM simulations, we used the same values of cosmological parameters as used by the TPM code, to generate the power spectrum by halofit. These parameters are, $\Om=0.3$, $\O_{\Lambda}=0.7$, $\O_{\mathrm{b}}=0.04$, dimensionless Hubble constant, $\h=0.7$, $\sigma_{8}=0.95$, and the derived power-spectrum shape parameter \citep{sugiyama_1995} $\Gamma \equiv \Om \h \exp\l[-\O_{\mathrm{b}}\l(1+\frac{\sqrt{2 \h}}{\Om}\r)\r]$.

\begin{figure}
\center
\epsscale{0.9}
\plotone{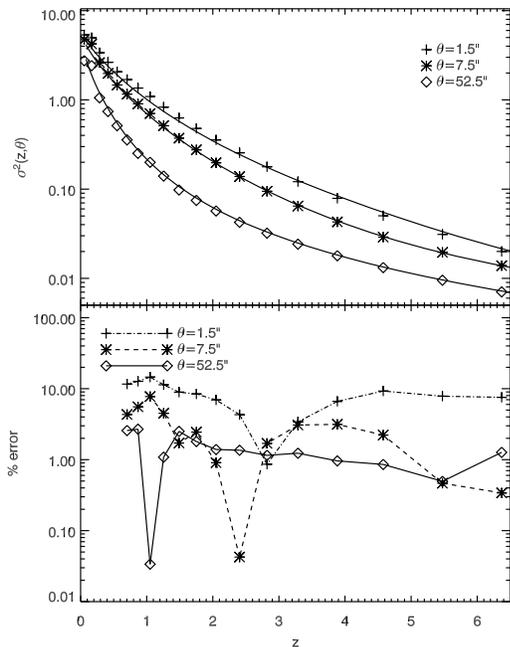}
\caption{\label{analvar} {\emph{Top}}: Comparison of the analytic result for the variance $\sigma^2_2(z,\theta)$ as a function of redshift $z$ for three different smoothing angles $\theta$, with the data obtained from the quoted TPM numerical simulations \citep{bode_ostriker_2003}. The points are the  variances from the N-body simulations. The lines represent the analytic result obtained by using the Limber equation and the nonlinear power spectrum of Smith et al. (2003). \emph{Bottom}: The percentage errors in the predicted values of the variance}
\end{figure}

We used redshifts corresponding to the center of each of  the $160\hmpc$ thick slices to generate the power spectrum for that slice. Then we used the Limber Approximation and the Gaussian filtering by the desired angle [\eqref{final}], to obtain the variances of the 2-D overdensities on the planes at each of these redshifts. The results are displayed in Fig.~\ref{analvar}. This test shows that we are able to predict the variance within each of these slices from theory to roughly $1-10\%$ accuracy.
\section{The Model}
\label{sec:model}
The probability distribution function for the lensing convergence for large angular scales, $> 10'$, can be straightforwardly obtained from analytic perturbative calculations \citep{stebbins_1996, villumsen_1996, bernardeau_etal_1997, jain_seljak_1997, kaiser_1998, schneider_etal_1998, vanwaerbeke_etal_1999}. However, on angular scales less that $5'$, the perturbative approach fails and  non-linear gravitational clustering comes into play. Hence, a number of models have been developed to move from a linear to the fully nonlinear power spectrum, and from there to the PDF of the convergence field. On intermediate scales $(1-10')$ such analytical models  have been shown to have good predictive power when compared with numerical simulations \citep[see][and references therein.]{barber_etal_2004}. {Recently, \cite{kayo_etal_2001} analyzed the one- and two-point statistics of the three dimensional dark matter distribution using $N$-body simulations and found them to be well described by the lognormal model \citep{kayo_etal_2001}. Following this, \cite{taruya_etal_2002} have shown that the lognormal model, with certain parameterizations, can fairly describe the convergence field for small angular scales $2'\la \theta \la  4'$ and up to a level threshold $\nu\sim 10$, where $\nu\equiv \k/\av{\k^2}^{1/2}$.  However, satisfactory analytic methods to generate the convergence PDF on the ultra non-linear scales, $<1'$, are as yet unknown.}  Here, we put forward a phenomenological model for the PDF of the surface mass density and hence, that of the convergence on the lens sheets, which seems to work down to  few arcseconds and is designed to better approximate the high density tail of the distribution, which reduces to a distribution akin to the lognormal and then to the Gaussian for large angles and high redshifts. The model is completely defined by the variance of the surface mass density on each sheet and thus encodes  all higher order statistical information in the two point statistics  of $\Sig$.\par
We computed the probability density function (PDF) of the quantity $x=\frac{\Sigma}{\av{\Sig}}\equiv 1+\delta_2$ on each plane. It is found that the PDF is well described by the following form, inspired by the lognormal distribution,
\begin{equation}
\label{mln}
f(x)=\frac{N}{x} \exp\l[-\frac{(\ln x+\omega^2/2)^2(1+ A/x)}{2 \omega^2}\r],
\end{equation}
where the values of the three parameters, $A$, $\o$ and $N$ are fixed by the following three constraints,
\beqn
\label{constr1}
\text{Normalization}:&\int_{0}^{\infty} f(x) dx& = 1,\\
\label{constr2}
\text{By defn. $\av{x}=1$ :}&\int_{0}^{\infty} x f(x) dx& = 1, \\
\label{constr3}
\text{Observed variance:}&\int_{0}^{\infty} (x-\av{x})^2 f dx=&\sigma_2^2. 
\eeqn
Therefore the model is fixed by a single input, the variance $\sigma^2_2(\theta,z)$ of the projected overdensity of at a given redshift and a given angular scale of smoothing. Since this quantity can be theoretically predicted, as discussed in the previous section, the model is essentially fixed by the cosmology.\par
A search algorithm is developed which, given a value of $\sigma_2^2$, returns the values of the three parameters satisfying all above constraints.\par

\begin{figure}
\centering
\epsscale{0.9}
\plotone{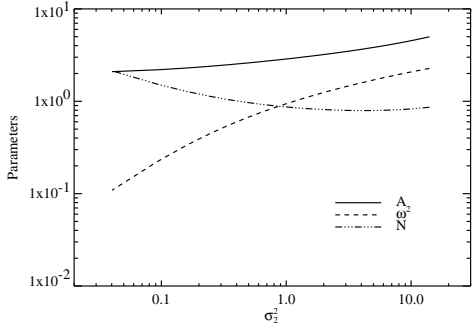}
\caption{The parameters $A$, $\o^2$ and $N$ plotted as a function of $\sigma_2^2(x)$ (cf. \eqref{mln})\label{params}}
\end{figure}

\begin{figure}
\centering
\epsscale{0.9}
\plotone{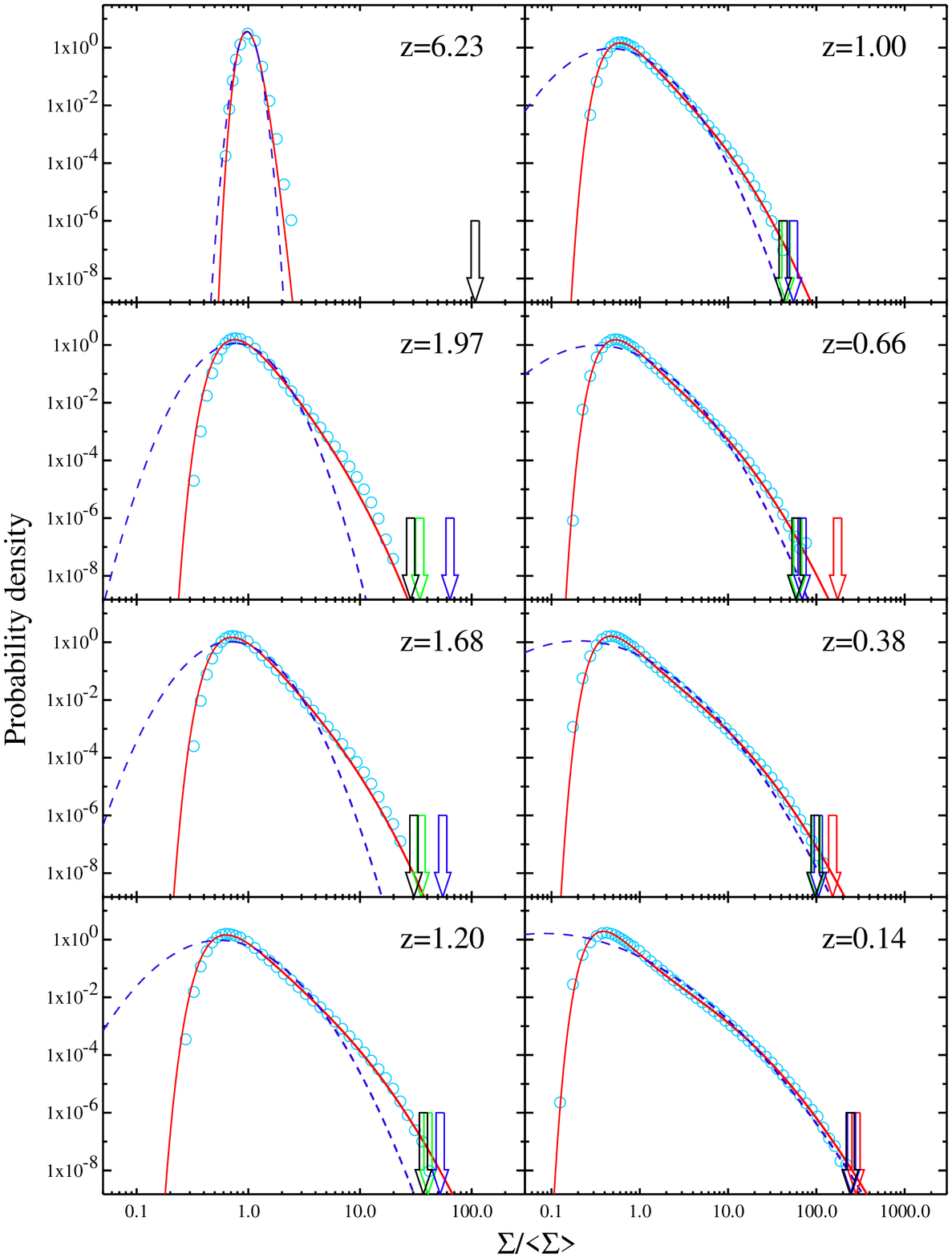}
\caption{The theoretical model (red line)  shown against the probability densities (blue circles) obtained from the quoted TPM simulation \citep{bode_ostriker_2003} for different redshifts. The dashed dark blue line is the lognormal model. The smoothing angle for this set of data was $7.5''$. The color coded arrows show the position of the critical surface density $\Scr$ in units of $\av{\Sig}$, for four different source redshifts: $7.0$ (black), $5.0$ (green), $3.0$ (blue) and $1.0$ (red). \label{fit05}}
\end{figure}

\begin{figure}
\centering
\epsscale{0.9}
\plotone{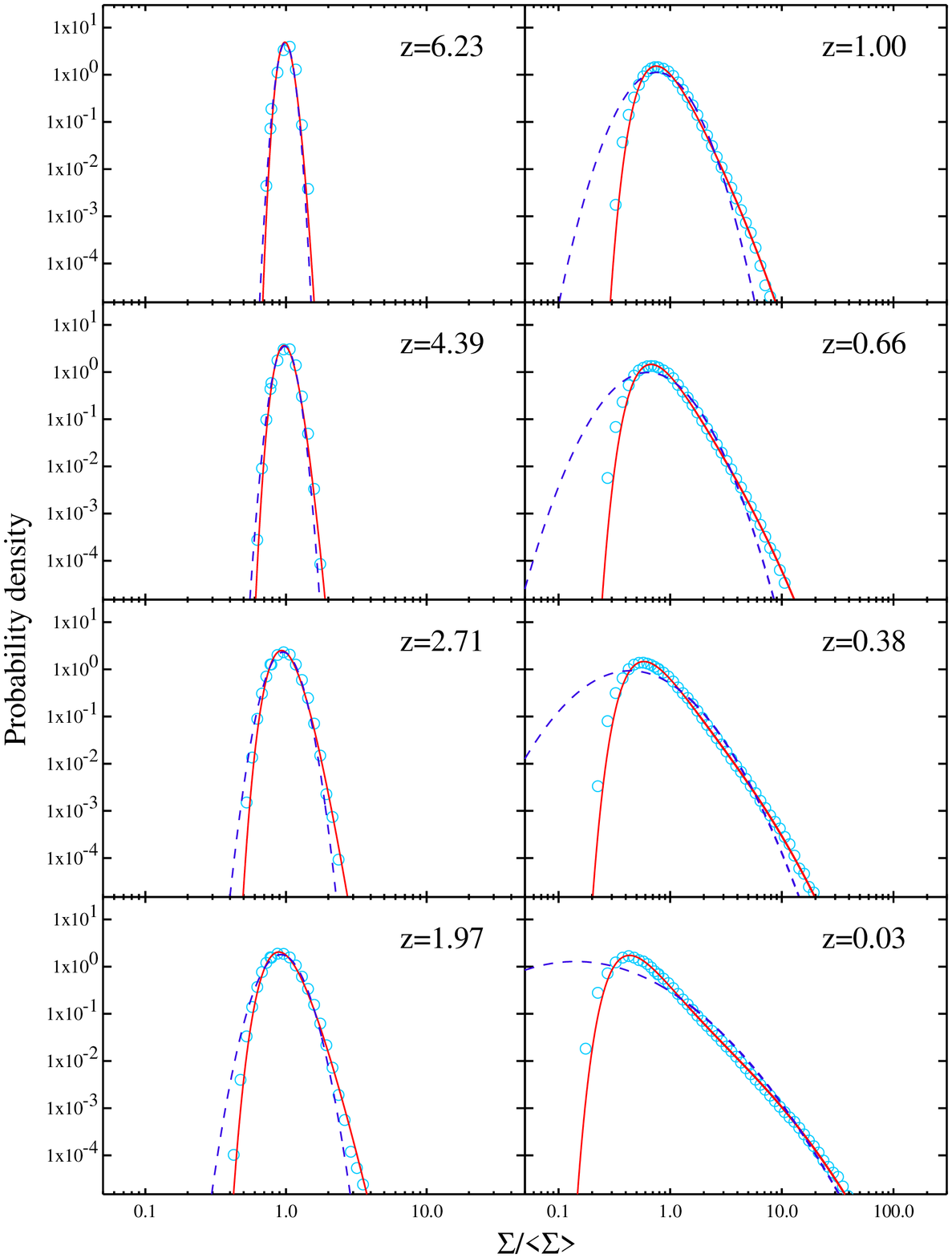}
\caption{\label{fit35} The theoretical model (red line)  shown against the  probability densities (blue circles) obtained from the quoted TPM simulation \citep{bode_ostriker_2003} for different redshifts.The dashed dark blue line is the lognormal model.  The smoothing angle for this set of data was $52.5''$. }
\end{figure}

\begin{figure}
\centering
\epsscale{0.9}
\plotone{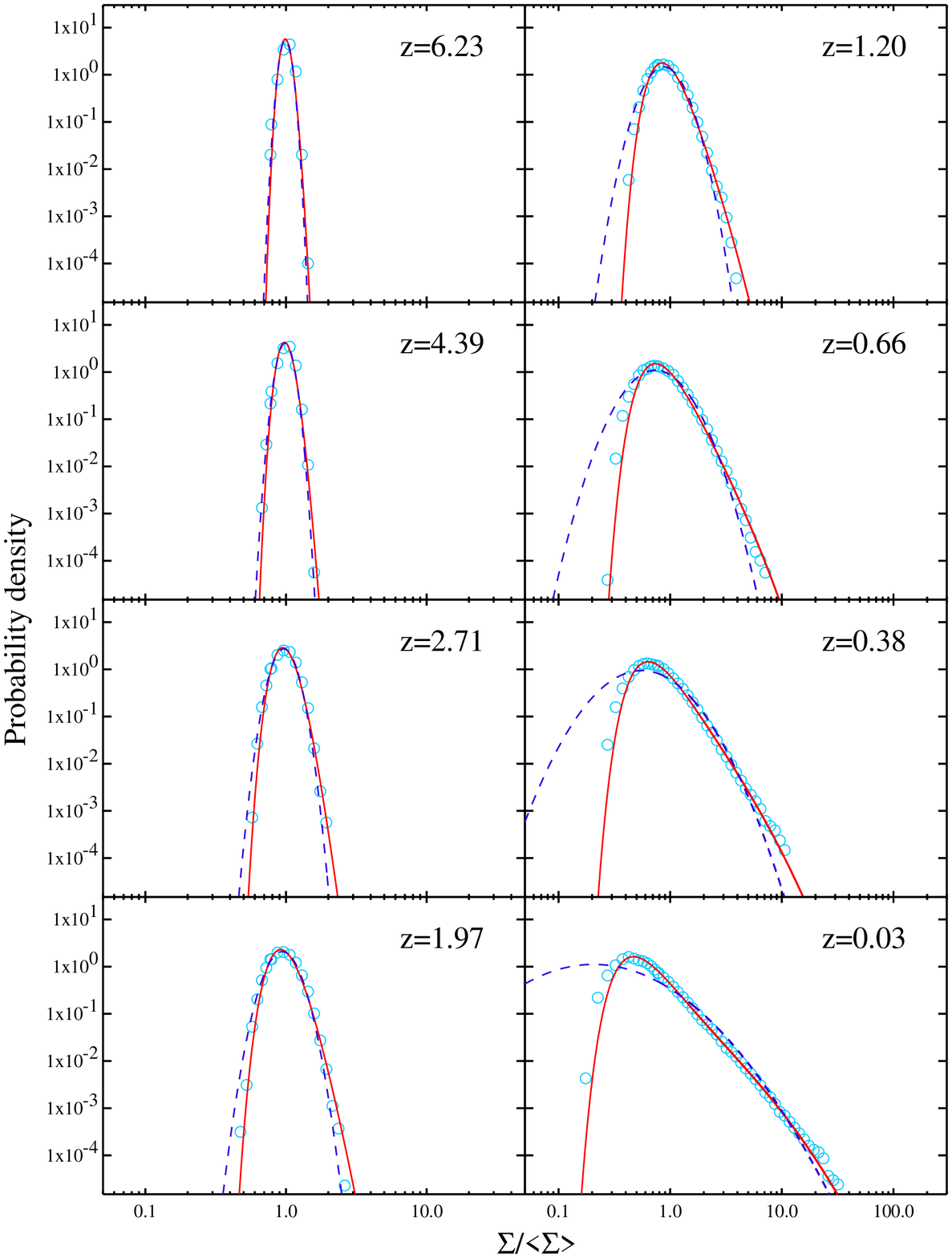}
\caption{The theoretical model (red line)  shown against the probability densities (blue circles) obtained from the quoted TPM simulation \citep{bode_ostriker_2003} for different redshifts. The dashed dark blue line is the lognormal model. The smoothing angle for this set of data was $90''$ or $1.5'$. \label{fit60}}
\end{figure}

\begin{figure}
\centering
\epsscale{0.9}
\plotone{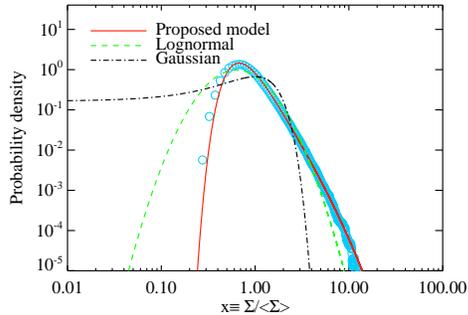}
\caption{Comparison of the lognormal and the Gaussian probability density functions with the model proposed in the paper. The data used for this illustration is for a redshift of $0.7$ and a smoothing angle of $52.5''$.\label{comp}}
\end{figure}

{The result is displayed in Fig.~\ref{params}, where the parameters $A$, $\o^2$ and $N$ are plotted as functions of $\sigma_2^2(x)$. Note that all these parameters are approximately power-laws of $\sigma_2^2$ for small values of the latter. \par 
}
 The theoretical model is shown against the PDF obtained from simulations for angular smoothing radii of $7.5''$, $52.5''$ and $1.5'$ in Figs.~\ref{fit05}, \ref{fit35} and {\ref{fit60}}, respectively, with  the lognormal model plotted over each. A detailed comparison of the model and other standard PDFs are displayed in Fig.~\ref{comp}.\par
 {We would like to stress here that although our model has three variables, $A$, $\omega$ and $N$, it is not a three parameter fit in the sense that the three variables are not varied to obtain some kind of ``best'' fit with the data points. We have simply  put forward a non-standard probability distribution which is on the same footing as the Gaussian or the lognormal as far as the number of free parameters is concerned. All the three distributions have their variables determined by the fact that they are normalized, have unit mean by construction and have a certain variance predicted by the cosmology. Thus, each of the three distributions is essentially fixed by the input cosmology and the knowledge of the redshift and the smoothing angle. In Figs.~\ref{fit05}, \ref{fit35} and {\ref{fit60}} we have simply overlaid the theoretical model on the data points; there is no fitting with the data involved anywhere.}

  It is immediately apparent from these figures that our model {does better} than the lognormal or the Gaussian in describing the PDF of the surface density, in particular it better describes the all important tail of the distribution which characterizes the low probability but observationally important high $\kappa$ events. The lognormal hugely underestimates the probability density for just those rare events in the tail of the distribution which are important for strong lensing. This effect is more pronounced for smaller smoothing angles, as evident from Fig.~\ref{fit05}, where we show, by arrows, the critical surface densities  required to produce strong lensing, for four representative source redshifts. At these values, the probability densities predicted by the lognormal model are orders of magnitude lower than those obtained from the TPM simulation. For all smoothing scales, it appears that on the right tail of the distribution, the lognormal does moderately well as long as probability densities are above $\sim 10^{-2}$, below which it departs from the data points.  Our model, however, appears to {describe} these heavy tails acceptably. Thus the lognormal proves to be a poor approximation for the very high surface density tails, except for very low or very high redshifts. In the latter limit, both of these models tend to the Gaussian, as will be shown below, and become almost indistinguishable. Another aspect of the distribution that the lognormal fails to describe completely is its skewness. The simulated distributions  fall sharply  toward the low surface density end, whereas the lognormal hugely overestimates the probability density in this region. {The proposed model appears to capture this feature of the distribution with appreciable accuracy}. It goes without saying that the Gaussian fails in all respects, except at  high redshifts and of course, for large enough smoothing angles for which the underlying Gaussian initial spectrum of perturbations remains dominant.  (cf. Fig.~\ref{comp}). 

\begin{figure}[h]
\epsscale{0.9}
\plotone{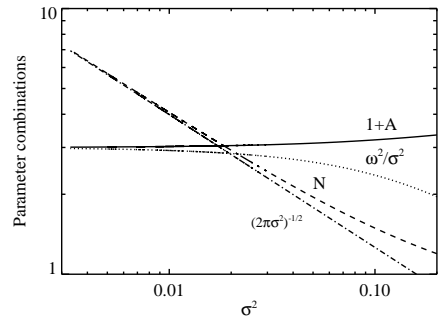}
\caption{At small $\sg^2$, the parameters $(1+A)$ and $\frac{\omega^2}{\sg^2}$ become identical implying $(1+a_0)=c_0\simeq 3$ (cf. \eqref{expand}). Also, the normalization parameter $N$ tends to the Gaussian value $1/{\sqrt{2\pi}\sg}$. These imply that the distribution tends to the Gaussian form with variance $\sg^2$ for small values of the latter.\label{smallsg}}
\end{figure}

Now, we shall discuss some limiting behavior of the above model. For compactness, we shall drop the subscript $2$ in $\sigma_2$ in the following discussion. Let us start by considering the exponent in the above formula, namely,
\beq
\label{arg}
-\frac{(\ln x+\omega^2/2)^2(1+ A/x)}{2 \omega^2}
\eeq
Since, $\o^2$ and $A$ appear to be well behaved functions of $\sigma^2$, one can write, 
\beqn
\label{expand}
\nonumber \o^2&=&c_0 \sigma^2 + O(\sigma^4)\\
A&=& a_0 + a_1 \sigma^2+O(\sigma^4) 
\eeqn
We can expand each component of \eqref{arg} as,
\beqn
(\ln x+\o^2/2)^2&=&(\ln x)^2+c_0 (\ln x) \sigma^2 + O(\sg^4)\\
1+\frac{A}{x} &=&1+\frac{a_0}{x}+\frac{a_1}{x}\sg^2+ O(\sg^4)
\eeqn
Therefore, the argument of the exponential can be expanded as,
\beqn
\nonumber
\frac{(\ln x+\omega^2/2)^2(1+ A/x)}{2 \omega^2}&=&\frac{1}{2 c_0 \sg^2}\left[\vphantom{\int}(\ln x)^2 {(1+\frac{a_0}{x})}\right.\\\nonumber
&+& \left\{ a_1 \frac{(\ln x)^2}{x}+c_0 (1+\frac{a_0}{x})\ln x\right\}\sg^2\\
&+&\left.O(\sg^4)\vphantom{\int}\right] 
\eeqn
For  $\sg^2\ll 1$ up to order $O(\sg^{-2})$, this reduces to,
\beqn
\frac{(\ln x+\omega^2/2)^2(1+ A/x)}{2 \omega^2}&=&\frac{1}{2 c_0 \sg^2}(\ln x)^2 {(1+\frac{a_0}{x})}
\eeqn
For small $\sigma^2$, the range of $x$ over which the exponent remains appreciable, is also small,
$$x=1+\delta$$ 
where $\delta^2 \sim O(\sigma^2)$.  In this approximation,
$$\ln x =\ln(1+\delta)\simeq \delta $$ and 
$$ \frac{1}{x}=1-\delta .$$
This implies,
\beqn
\frac{(\ln x+\omega^2/2)^2(1+ A/x)}{2 \omega^2}&\simeq&\frac{\delta^2}{2 c_0 \sg^2}{(1+{a_0})}
\eeqn
Under these approximations the model reduces to, 
\beqn
\nonumber
\frac{N}{x} \exp\l[-\frac{(\ln x+\omega^2/2)^2(1+ A/x)}{2 \omega^2}\r]&\simeq& {N} \exp\l[-\frac{\delta^2}{2 c_0 \sg^2}{(1+{a_0})}\r]\\
&=& {N} \exp\l[-\frac{\delta^2}{2 \beta^2\sg^2}\r]\eeqn
Where we have defined $\beta\equiv(1+a_0)/c_0 $. One can easily verify that the condition in \eqref{constr3} gives, $\sigma^2=\sigma^2 \alpha^2$, implying $\alpha=1$. So the normalization condition, \eqref{constr1} gives us $N=1/(\sqrt{2 \pi } \sigma)$. This is also graphically illustrated in Fig.~\ref{smallsg}. Thus our model correctly reduces to a Gaussian for small values of variance. 

For small $\sg$, we have $1+a_o=c_0\simeq 3$ and therefore,
$$\o^2\simeq 3\sg^2+O(\sg^4)$$
At this point, it is important to note that the surface density, $\Sig$, that we have been considering so far, contains in it the contribution from the smooth background matter density $\av{\Sig}$. However, lensing is sensitive only to the departures in the surface mass density from the average rather than its total value {\citep[equations 3.32 and 3.48]{petters_levine_wambsganss}. One should note in this connection that the values of $\Sig$ relevant to strong lensing events are usually much higher than the background density $\av{\Sig}$. This often leads to neglecting the subtraction of background density in the definition of  $\kappa$ when discussing strong lensing. In many cases, this is indeed a very small effect. However, we have found appreciable differences, especially in the multiple plane lensing probabilities (see Section \ref{sec:prob}), if this correction is not taken into account. Therefore, we define the lensing surface mass density $\Sig_l$, that usually appears in the lensing equations as,}
\beqn
\label{sigmalens}
\Sig_l\equiv\Sig-\av{\Sig}.
\eeqn 
For the purpose of studying the possibility of strong lensing by a plane, it is important to notice that a multiple image of the source results if the beam encounters a background-subtracted surface density, $\Sig_l$ along its path which is in excess of the critical surface mass density $\Scr(z)$ at that redshift, which is given by, 
\beq
\Scr(z)=\frac{c^2}{4 \pi G} \frac{D_s}{D_{ls} D_l}, 
\eeq
where $D_s$, $D_l$ and $D_{ls}$ are the angular diameter distances from the observer to the source, the observer to the lens, and the lens to the source, respectively. \par
In order to address the problem in which one, two or more planes together become supercritical, it is convenient to change the variable of interest from $x=\frac{\Sig}{\av{\Sig}}$ to the convergence,
\beq
\label{kappa}
 \kappa=\frac{\Sig_l}{\Scr}={\av{\Sig}\over{\Scr}} (x-1).
\eeq
The PDF $g(\kappa)$ for the convergence is related to that of $x$, $f(x)$  by a simple Jacobian transformation, 
\beq g(\k)=\alpha f(\alpha \k+1),\eeq
with, $\alpha={{\Scr}\over{\av{\Sig}}}.$\par
{A physical meaning to the variable $\alpha$ can be attached by defining the minimum value $\kmin$ of the convergence $\kappa$, corresponding to an ``empty'' beam, with $\Sig=0$. In this case, \eqref{sigmalens} implies $\Sig_l=-\av{\Sig}$ and hence, }
\beq
\label{kmin}
\kmin=-\frac{\av{\Sig}}{\Scr}=-\frac{1}{\alpha}.
\eeq
Therefore, one may write $\alpha=\frac{1}{|\kmin|}$ and the PDF for the convergence can be rewritten as, 
 
\beq g(\k)=\frac{1}{|\kmin|} f\l(\frac{\k}{|\kmin|}+1\r),\eeq
or, written out in its entirety as,
\beqn
\label{kmln}
\nonumber g(\k)d\k=N&&\\
\nonumber\times& \exp\l[-\frac{(\ln (1+\k/|\kmin|)+\omega^2/2)^2\{1+ A/(1+\k/|\kmin|)\}}{2 \omega^2}\r]&\\
 \times &{\frac{d\k}{\k+|\kmin|}}.\hphantom{this is just to push dkappa/kappa}&
\eeqn
The constraint equations, (\ref{constr1}), (\ref{constr2}) and (\ref{constr3}), translate to, 
\beqn
\label{kconstr1}
\text{Normalization}:&\int_{\kmin}^{\infty} g(\k) d\k& = 1,\\
\label{kconstr2}
\text{By defn. $\av{\k}=0$ :}&\int_{\kmin}^{\infty} \k g(\k) d\k& = 0, \\
\label{kconstr3}
\text{Observed variance:}&\int_{\kmin}^{\infty} \k^2 g(k) d\k&=\av{\k^2}. 
\eeqn
This transformation leads us to the possibility of expanding the applicability of our model beyond the thin mass-sheet case that we have been considering so far. {Although we will not consider this here, it may be interesting to ask how much improvement this model provides over the lognormal form studied by \cite{taruya_etal_2002} in describing the line-of-sight or effective convergence, given by \citep[see, e.g.][]{bartelmann_schneider_2001},
\beq
\label{keff}
\k_e=\frac{3}{2}\Om\l(\frac{H_0}{c}\r)^2\int_0^{\chi_s} d\chi g(\chi,\chi_s)\delta(\chi),
\eeq
 where $\chi$ is the comoving distance, $\chi_s$ is the comoving distance to the source, $\delta=\Delta \rho/\av{\rho}$ is the density contrast, and
\beq
g(\chi,\chi_s) = r(\chi)\frac{r(\chi_s-\chi)}{r(\chi_s)} (1+z),
\eeq
 $r(\chi)$ being the comoving angular diameter distance. This will be the subject of a future paper. }
\section{ Strong lensing probabilities }\label{sec:prob}
{The necessary condition for a strong lensing situation is $\k_e\simeq \sum \k_i>1$, where $i$ is the index of a plane in front of the source. As the convergence $\k$ at a point on each plane can be either positive or negative, a supercritical plane does not necessarily imply a supercritical line of sight . {In fact, as \cite{premadi_martel_2004} note, underdense regions tend to occur around cluster lenses}. However, when a single plane or a number of planes, collectively, is supercritical, the probability that negative convergences on some of the remaining planes will make the line of sight subcritical is actually small.  Neglecting this real but small correction for the moment, we shall show that there exists a possible analytical calculation to compute the probability that a single plane produces strong lensing. Then, in a following subsection we shall turn to a more accurate numerical (Monte Carlo) approach.} \par
Notation: Let $\cf_i$ denote the probability that the $i$-th plane is supercritical, i.e. 
\beq\label{fi}{\cf}_i\equiv{\text{Prob}}\left[\Sigma(z_i)>\Scr(z_i)\right]=\int_{1}^{\infty} g(\kappa(z_i)) \d\k(z_i).\eeq
\subsection{Probability that one and only one plane is supercritical}
This is equal to the probability that the first plane is supercritical and all the other planes are not, {\bf or} the second plane is supercritical and the other planes are not , and so on: 
\beqn
\label{pone}
{\cal P}_{\mathrm{one}}&=&\sum_{i=1}^{N} \cf_i \prod_{\scriptsize\begin{array}{c} j=1\\j\ne i\end{array}}^{N}(1-\cf_j)\nonumber\\
&\simeq& \sum_{i=1}^{N} \cf_i (1-\sum_{j\ne i} \cf_j),
\eeqn
where $N$ is the number of planes in front of the source, and the last step follows from the fact that $\cf_i\ll 1, \forall i$. An extension of this scheme for  two or more planes becomes increasingly cumbersome and impractical as the number of planes increases. Also, as we already pointed out, this method does not correct for the fact that some lines of sight may become subcritical due to negative convergences on the planes other than those considered in the probability calculation. Therefore, we turn to the more efficient Monte Carlo approach to study these probabilities. 
\subsection{Monte Carlo method.}
We randomly draw values, $\k_i$, from the PDFs of $\kappa$ on the planes lying in front of the source and consider only those cases for which the total convergence is greater than unity. With $\sum \kappa_i>1$ satisfied, if the highest value among the $\kappa_i$'s is greater than unity we identify this as a single plane strong lensing case. If this is fails, we consider the two highest values of $\kappa$ from our crop and look at the sum of the two. If the sum exceeds unity then we identify this as a two plane lensing case. We extend this method to include three or more planes. 
\begin{figure}
\center
\epsscale{0.90}
\plotone{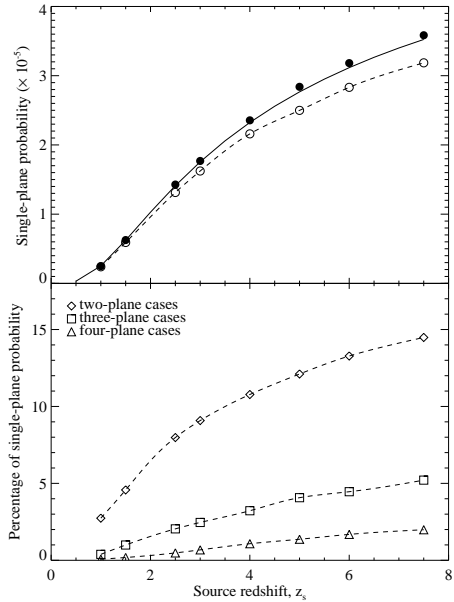}
\caption{\emph{Top:} The probability, as a function of source redshift, that a single plane alone renders the line of sight supercritical. The smoothing angle for this figure was $7.5''$.  The solid line is the analytical approximation (cf. \eqref{pone}). The filled circles represent the single-plane probabilities generated by a Monte Carlo simulation which does not take into account the correction discussed in the text. The open circles represent the same, but with the correction applied. As discussed in the text, the analytical calculation slightly overestimates the true probabilities (open circles) as it includes rare cases for which, although the single plane is supercritical, the line of sight is rendered subcritical by negative convergence on some other plane or planes. The open circles are joined by a spline (dashed line) as a guide to the eye. {\emph{Bottom:}} The probabilities, shown as the percentage of the corresponding single-plane values, that two, three, or four planes together render the line of sight supercritical. The points are from the Monte Carlo study and are joined by splines (dashed lines) to aide the eye.  \label{oneplane}}
\end{figure} 
\subsection{Results}
{As we have already shown, the probabilities for the single-plane cases are easy to compute analytically (cf. \eqref{pone}), as long as we do not care whether underdensities on the other planes are making the line of sight subcritical. {With this assumption, the analytical calculation slightly overestimates the probabilities as expected. However, it is useful as a quick and handy way to get an upper limit to the probabilities. The results from the Monte Carlo study do take the effect of underdense planes into account.} The results from both the analytical and the Monte Carlo approach, for a smoothing angle of $7.5''$, are displayed in Fig~\ref{oneplane}. We note from the figure that when the constraint $\sum \kappa >1$ is not applied in the Monte Carlo study and cases with $\kappa_{\mathrm{max}}>1$ but $\sum \kappa <1$ are wrongly counted as legitimate single-plane cases, the single-plane probabilities obtained analytically match those from the Monte Carlo simulation, as expected. When the correction is taken into account, the points from the Monte Carlo study lie slightly below the analytical values, because we lose a few cases due to the fact that the line of sight was subcritical although the single plane was supercritical.  We notice that for source redshifts, $z_s \simeq 1.0$, the contribution of secondary clumps of matter to strong lensing is almost negligible ($\sim 2-3\%$ of the single-plane cases). However, for larger redshifts secondary planes become more and more important and contribute up to about $15\%$ of the single-plane cases for $z_s=6.0$. Our results are somewhat lower than those obtained in a previous study by \cite{WBO_2004b}. This may be due to various reasons, including the fact that we have used a smoothed surface density distribution and are considering a larger opening angle than was done in \cite{WBO_2004b} . However, our results seem to be in good agreement with those quoted in \cite{dalal_hennawi_bode_2005}}, who report a secondary plane contribution of $\la 7\%$ for a source redshift, $z_s=2.0$ and angular scales $\ga 10''$.

\section{Error in the mass estimation of the main lens}\label{sec:mass}
\begin{figure}
\centering
\epsscale{0.9}
\plotone{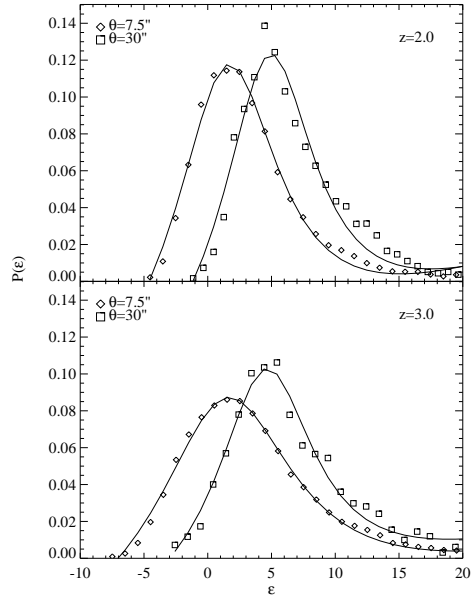}
\caption{Probability distribution of the percentage error, $\epsilon$, in the mass of the main lens for two different smoothing angles. {\emph{Top}}: Source redshift, $z_s=2.0$. {\emph{Bottom}}: Source redshift, $z_s=3.0$.   \label{foreground}}
\end{figure}

As discussed in the previous section, there is a finite and non-negligible probability that more than one lens is responsible for causing a strong lensing event. We may ask, by how much we can underestimate the mass in lensing cluster or galaxy by assuming that it is the only object responsible for lensing. We study this by again shooting rays through the planes and considering cases where the total convergence along the line of sight,  $\kappa_{\mathrm{tot}}\equiv\sum \kappa_i$ is greater than unity. Within these cases, we look at the difference between $\kappa_{\mathrm{tot}}$ and the maximum value, $\kappa_{\mathrm{max}}$ among the $\kappa_i$'s on the  planes in front of the source. This maximum value would be the true value of the convergence due to the primary cluster which is the major contributor to the lensing. If we think that the other planes do not contribute significantly, we would assign the value $\kappa_{\mathrm{tot}}$ to it instead of $\kappa_{\mathrm{max}}$ and this will be an error because statistically the contribution due to secondary planes is not negligible. To obtain an estimate of the error, we  repeat the exercise several times and draw up the probability distribution of the percentage error in $\kappa$, defined as,
\beq
\epsilon=\frac{\kappa_{\mathrm{tot}}-\kappa_{\mathrm{max}}}{\kappa_{\mathrm{max}}}\times 100.
\eeq  
We do this for a source redshifts of $2.0$ and $3.0$ and for each redshift, draw up two probability distributions of $\epsilon$ corresponding to the smoothing scales of $7.5''$ and $30''$. The results are displayed in Fig.~\ref{foreground}. The distributions for the smaller smoothing length $(\sim 7.5'')$ has a much longer tail (not shown),  extending to very high values of $\epsilon$, than that  for the larger smoothing angle. This is expected because the probability density function of $\kappa$ has a heavier tail for small smoothing angles, than it has for larger ones. We have also attempted to quantify the central value and the dispersion in $\epsilon$ by fitting a Lorentzian-like function to the data, given by,

\beq
\label{lorentzian}
f(x)=\frac{a_0}{[(x-a_1)/a_2]^2+1}+a_3 x,
\eeq
where the best fit values of the parameters $a_1$ and $a_2$ are interpreted as the centroid and the dispersion (half-width at half-maximum, or HWHM) of the distribution. These values are tabulated in Table~\ref{fitlorentz}.
\begin{deluxetable}{cccccc}
\tablewidth{0pc}
\tablecolumns{6}
\tablecaption{Parameters obtained by fitting the formula in \eqref{lorentzian} to the pdf of $\epsilon$.  \label{fitlorentz}}
  \tablehead{ Source redshift ($z_s$) &\multicolumn{2}{c}{Peak centroid ($a_1$)} &\colhead{}&\multicolumn{2}{c}{ HWHM ($a_2$)}\\
    \cline{2-3}\cline{5-6}\\
   \colhead{} & \colhead{$\theta=5''$} & \colhead{$\theta=20''$}&\colhead{} & \colhead{$\theta=5''$} &\colhead{$\theta=20''$} }
    \startdata
    2.0&1.52 & 4.82 & &4.93 & 4.39 \\
    3.0 & 1.45 & 4.55 && 6.52 & 4.62 \\
\enddata
\end{deluxetable}
\par
{From the table we note that the dispersions in the values of $\epsilon$ around the peak increases with increasing source redshift for a fixed smoothing angle. This is expected because for higher redshifts there are more intervening planes and hence a larger range of masses in clumps lining up with the main lens. Also, at a given redshift, the spread in the values of $\epsilon$ is larger for a smaller smoothing angle. This, basically reflects the fact that  the variance of $\kappa$ is higher for smaller smoothing angles. We also note, that although the distributions for the smaller smoothing angle has a much longer tail (not shown), the most probable error corresponding to it is lower than that for the larger smoothing angle. This basically means that although multiple-plane cases are less probable for the larger smoothing angle, when they happen, they carry a larger contribution of mass from the auxiliary planes, than in a typical small angle case.  For instance, in a two-plane case the value of $\k$ on the secondary plane typically come from the region around the peak of the PDF of $\k$ on that plane. For a small smoothing angle, the peak is close to $\k\la 0$ (the ray mostly hits slightly underdense regions), whereas as the smoothing angle grows the peak moves towards more positive values of $\k$. So in a  supercritical line of sight, secondary planes will typically contribute higher values of $\kappa$ when the smoothing angle is larger. This also applies to all the planes along the line of sight in general, so that for a larger smoothing angle the residual sum from all other planes, $\ktot-\kmax$, will be higher than that for a smaller smoothing angle.  }

\par
\section{Effect of inhomogeneities on angular diameter distances}\label{sec:angdia}
{Inhomogeneities in the distribution of matter induce fluctuations in the measured value of the angular diameter distance at a fixed redshift, z. The problem has been attacked directly from general relativity \citep{zeldovich_1964, refsdal_1970, dyer_roeder_1974,sasaki_1987, futamase_sasaki_1989,watanabe_tomita_1990, segal_1993, kantowski_etal_1995, linder_1998, kantowski_etal_2000, mortsell_2002,demianski_etal_2003}  producing results of varied complexity.} \par   
 In the following, we shall show that under certain simplifying assumptions, the differential equation for angular diameter distances can be obtained as a continuum limit to multiple gravitational lensing by successive mass-sheets. This will provide a framework for studying angular-diameter distances in an inhomogeneous universe as a multiple deflection problem. The calculations are much in the spirit of \cite{hadrovic_binney_1997}, although we have generalized their results to include a non-zero cosmological constant. \par
Let us use the convention that $\ti D$ and $D$, respectively, denote
angular-diameter distance before (empty beam) and after lensing is taken into account.

Fig.~\ref{lenses} shows a light ray that passes two mass sheets at impact parameters (measured with respect to the fiducial ray, OO$'$) $\bxi_1$ and $\bxi_2$ and  is deflected through angles $\bm{\alpha}_1$ and $\ba_2$, respectively. A simple geometrical argument \citep[see, for e.g.][]{hadrovic_binney_1997} shows that the impact parameter, $\bxi_j$ on the $j$-th plane is related to $\bxi_1$ through the relation,
 \beq\label{lens_seq}
 \bxi_j={{\ti D_j}\over{\ti D_1}}\bxi_1
-\sum_{i=1}^{j-1}{\ti D_{ij}\ba_i(\bxi_i)},
\eeq
{where ${\ti D_i}$ is the empty-beam angular diameter distance to the $i$-th plane and $\ti D_{ij}$ denotes the empty-beam angular diameter distance between the $i$-th and the $j$-th planes.}\par
If we assume that the mass distribution inside the beam on each mass sheet is uniform, with a surface  density $\Sigma$, then the angle of deflection is given by,
 \beq
\ba={4\pi G \over {c^2}}\Sigma\bxi.
\eeq
 Hence in this case the impact parameters are all parallel and satisfy
\beq\label{all_xis}
 \bxi_j={{\ti D_j}\over{\ti D_1}}\bxi_1
-{4\pi G \over {c^2}}\sum_{i=1}^{j-1}\Sigma_i\ti D_{ij}\bxi_i.
\eeq
 The angular diameter distance to an object with diameter $\xi_i=|\bxi_i|$ is
 $D_i\equiv\xi_i/(\xi_1/\ti D_1)$. Therefore, on taking the
scalar form of \eqref{all_xis} and dividing throughout by $(\xi_1/\ti D_1)$ we find that the equation satisfied by the true angular diameter distances $D_j$ has the form, 
\beq\label{disc_ang_diam}
 D_j=\ti D_j-{4\pi G \over {c^2}}\sum_{i=1}^{j-1}\Sigma_iD_i\ti D_{ij}.
\eeq
{\begin{figure}
\epsscale{0.9}
\plotone{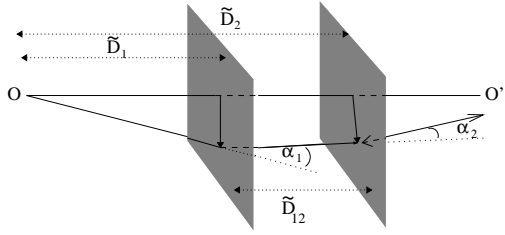}
\caption{Lensing by a series of mass sheets.\label{lenses}}
\end{figure}}
\subsection{Equation for $\ti D(y,z)$.} 
{Let us denote the redshifts of the $i$-th and the $j$-th planes as $y$ and $z$, respectively. Also, let the continuous version of the distance $\ti D_{ij}$  be denoted by $\ti D(y,z)$.}\par
$\ti D(y,z)$ is the angular diameter distance in an empty beam (vanishing Ricci tensor and shear) and hence the focusing theorem {\citep[][equation (3.64)]{SEF_1992}} takes the form,
 \beq\label{focus_e}
{\d^2\ti D\over\d\tau^2}=0,
\eeq
where $\tau$ is the affine parameter, defined through,
\beq\label{tau}\d \tau={{\d z}\over{(1+z)^2\sqrt{\Omega_m (1+z)^3 +\Omega_k (1+z)^2+\Omega_{\Lambda}}}}\equiv {\d z\over{g(z)}},\eeq
{where we have defined,
$$g(z)\equiv {(1+z)^2\sqrt{\Omega_m (1+z)^3 +\Omega_k (1+z)^2+\Omega_{\Lambda}}}.$$}
With the help of \eqref{tau}, \eqref{focus_e}  can be expressed in terms of redshift as, 
\beq\label{D_empty}
g(z){\pa \over {\pa z}}\l( g(z) {\pa \over {\pa z}} \ti D(y,z)\r)=0.
\eeq
 The initial condition is given by \citep[][equation~(4.53), { same arguments, but with $\O_{\Lambda}\neq 0$}]{SEF_1992},
\beq\label{D_empty2}{\pa\ti D(y,z)\over\pa z}\bigg|_{z=y}={(1+y)\over g(y)}.
\eeq
\subsection{Transition to continuum}
Now we shall fill the beam with a cosmological matter density  and use equation (\ref{disc_ang_diam}) to calculate the angular-diameter distance to a redshift $z$. We consider matter in the beam to be sliced up into discs of infinitesimal thickness. {The disc that lies between redshifts $y+\d y$ and $y$ has surface density}
 \beqn \label{dsigma} \d\Sigma&=&\rho_m(y){\d r_{\rmn{prop}} \over \d y} \d y, \eeqn
{where, the proper distance, $r_{\mathrm{prop}}$ is given by,
\beqn\label{rprop}{\d r_{\rmn{prop}}}&=&-c \d t\nonumber\\
&=&{c \over{H_0}}{ {\d y}\over{(1+y)\sqrt{\Omega_m (1+y)^3 +\Omega_k (1+y)^2+\Omega_{\Lambda}}}}\nonumber\\
&=&{c \over{H_0}}\frac{(1+y)}{g(y)} 
\eeqn
and the physical density at redshift $y$, $\rho_m(y)$ is given by, 
\beq\label{rhom}
\rho_m(y)=\frac{3 H_0^2}{8\pi G} \O_m[1+\delta(y)] (1+y)^3, 
\eeq
where $\O_m$ is the dimensionless density parameter in the present universe and we have entertained the possibility of having inhomogeneous distribution of matter through the perturbation $\delta(y)$ in density.}\par
{Using equations (\ref{rprop}), (\ref{rhom}) and (\ref{dsigma}), we have,
\beqn
\label{sigma_emp}
\d\Sigma &=&\rho_m [1+\delta(y)](1+y)^3 {c \over {H_0}} {{(1+y)}\over g(y)} \d y\nonumber\\
&=&{ {3 c H_0} \over{8 \pi G}}\l[\O_m \l\{1+ \delta(y)\r\} {(1+y)^4 \over g(y)} \r]\d y. 
\eeqn}
{The continuum limit of the multiple deflection equation (\ref{disc_ang_diam}), after making the replacements: $\ti D_j\rightarrow \ti D(z)$, $D_i \rightarrow D(y)$, $\ti D_{ij}\rightarrow \ti D(y,z)$ and $\sum_{i=1}^{j-1} \rightarrow \int_0^z \d y$, takes the form,
\beqn
\nonumber D(z)&=&\ti D(z)- \frac{4\pi G}{c^2}\int \d y \frac{\d \Sigma}{\d y} D(y)\ti D(y,z).\\
\eeqn
Using \eqref{sigma_emp} and expressing  all distances in units of $c/H_0$, the above equation becomes,  
\beqn\label{d_int}
D(z)&=&\ti D(z)\nonumber\\
&&-\fracj32\Omega_m\int_0^z\d y {{[1+\delta(y)] (1+y)^4}\over g(y)}D(y)\ti D(y,z).
\eeqn}
We can convert this integral equation to a differential one as follows. Upon differentiating \eqref{d_int} with respect to $z$, we have,
 \begin{eqnarray}
{\d D\over\d z}&=&{\d\ti D\over\d z}\nonumber\\&&-\fracj32\,\O_m \,
\int_0^z\d y\,[1+\delta(y)]{(1+y)^4\over g(y)}D(y){\pa\over\pa z}\ti D(y,z),
	\nonumber
\eeqn
and writing $\O_m [1+\delta(y)]$ as $\Om(y)$ for brevity, we further have,
\beqn
{{\d  \over \d z}\l(g{\d D\over\d z}\r)}&=&{{\d \over \d z}\l(g(z){\d \ti D\over\d z}\r)}\nonumber\\
&&-\fracj32\,
\O_m(z)\,{(1+z)^4}D(z)\l({\pa\over\pa z}\ti D(y,z)\r)_{y=z}\nonumber\\ 
&&-\fracj32\,\O_m\,\int_0^z\d y\,\l[\{1+\delta(y)\}{(1+y)^4\over g(y)}D(y)\r.\nonumber\\
&& \hspace{2cm} \times \l.{ {{\pa}\over{\pa z}}\l( g(z){\pa\over\pa z}\ti D(y,z)\r)}\r]\nonumber.
\end{eqnarray}
The first and the last terms on the right hand side of the above equation vanishes by virtue of (\ref{D_empty}). Using (\ref{D_empty2}), we finally have the differential equations for $D(z)$ as,
\beq\label{dr}
g(z){{\d  \over \d z}\l(g(z){\d D\over\d z}\r)}+\fracj32\,
\O_m(1+\delta(z))\,{(1+z)^5}D(z)=0.
\eeq
If $\delta\equiv 0$ then \eqref{dr} is the correct differential equation for angular diameter distance in an FRW universe. If the ray propagates through a smooth under-dense region having a constant comoving density $\alpha \O_m$ ($0<\alpha<1$), then the equation reduces to the historical Dyer-Roeder equation.\par
Hence, we have shown that the angular diameter distances can be studied as multiple deflection lensing problem. It is important to discuss the simplifying assumptions that lead to this derivation. Firstly, we have assumed that the focusing of rays propagating over cosmological distances is unaffected by shear due to clumps of matter lying outside the beam. \cite{futamase_sasaki_1989} and \cite{watanabe_sasaki_1990} show that shear contributes negligibly to focusing as long as the scale of inhomogeneities are greater than or comparable to galactic scales. Since we study this problem with smoothing scales $\sim 10$\arcsec, this assumptions seems to be a reasonable one. Secondly, we have assumed that matter that lies between the redshifts $z$ and $z+\d z$ forms a uniform disc. This is a nontrivial approximation but one that is particularly suited to our setup, in which we are looking at surface mass distributions on planes which have been already smoothed by a window of a certain angular size.
\subsection{Implementation and results}
{\begin{figure}[t]
\epsscale{0.9}
\plotone{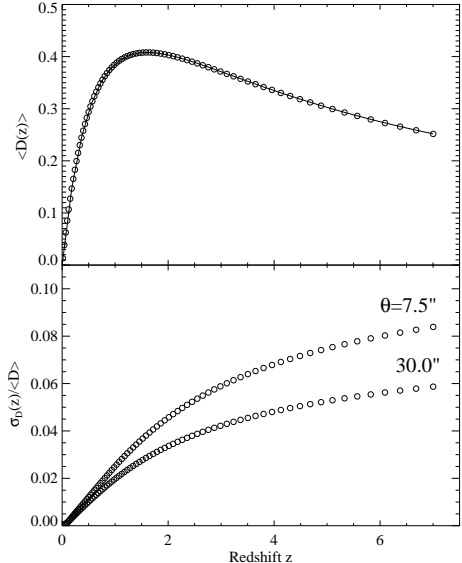}
\caption{\emph{Top}:~The mean angular diameter distances obtained by Monte Carlo sampling of surface densities from the lens planes. The continuous line represents the smooth Friedman Robertson Walker (FRW) angular diameter distance as a function of redshift. The data are for an angular smoothing scale of $7.5''$. \emph{Bottom}: The ratio, ${\sigma_{\scriptstyle D}(z)}/{\av{D(z)}}$, of the standard deviation to the mean of the angular diameter distance to each redshift, for two angular smoothing scales, $7.5''$ and $30.0''$.  \label{angdia}}
\end{figure}}
 For each beam, we approximate the continuous distribution with a Monte Carlo approach utilizing $38$ values of $\Sig_{\theta}$ from the PDFs corresponding to the $38$ planes and compute the angular diameter distances out to various redshifts using \eqref{disc_ang_diam}. In the top panel Fig.~\ref{angdia}, we show the result of a Monte Carlo run with $3\times 10^4$ beams, each having an opening angle of $7.5''$. As expected, the mean angular diameter distances perfectly match the smooth Friedman Robertson Walker (FRW) values. However, as  displayed in the bottom panel of the same figure, the ratio of the standard deviation to the mean value of the angular diameter distance, increases with increasing redshift, amounting to a probable error of  $\sim 5\%$ at a redshift of $\sim 2.0$ and rising up to $\sim 9\%$ for a redshift $\sim 7.0$, for the smoothing angle of $7.5''$ . The dispersions diminish with an increase in the smoothing angle, as is evident from the data points for the smoothing angle of $30''$.  This is expected because with a larger smoothing angle, the surface densities encountered by the beam departs from the mean density with lesser probability. In Fig~\ref{angdiahis}, we show the probability distributions of the angular diameter distance $D$, at four redshifts. These distributions are skewed towards values of $D$ lower than the FRW values and the skewness is higher for the smaller smoothing angle. The skewness implies that a typical ray more frequency passes through an underdense region than it does an overdense one, and as discussed before in relation to the error in lens mass estimation, this bias becomes less important as the smoothing angle increases.  Since various combinations of the angular diameter distance enter the calculations of lensing, the dispersion in its value, as displayed here, may systematically affect the use of lensing to determine of the properties of lenses or aspects of the underlying cosmology. 
{\begin{figure}[t]
\epsscale{0.9}
\plotone{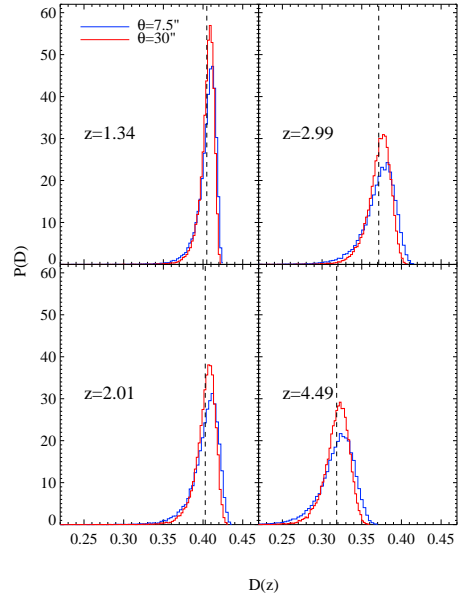}
\caption{The probability distribution function of the angular diameter distance to four different redshifts, $z=1.34,\: 2.01,\: 2.99$ and $4.49$ and for two smoothing angles $\theta=7.5''$ (blue histograms) and $30''$ (red histograms). The dashed vertical lines mark the position of the average or FRW angular diameter distance at each redshift.  \label{angdiahis}}
\end{figure}}
\section{Conclusions}
In the present paper, we have studied gravitational lensing with a multiple lens plane approach and have proposed a new analytic model for the dark matter convergence on the lens planes. Like the lognormal, this model is entirely defined by $\av{\k^2}_{\theta}$, the variance of $\k$ on a given lens plane and for a given smoothing angle $\theta$. We showed that these variances can be theoretically predicted from the knowledge of the nonlinear power spectrum under a given cosmological model. Our tests indicate that the proposed model is far better than the lognormal in the ultra-nonlinear regime of small smoothing angles ($< 1'$) and describes the low-probability, high $\k$ tail of the PDF of $\k$ with quite high accuracy. We illustrated a number of possible applications of the model. With the PDF of kappa on each plane accurately defined by our model, we could mimic an expensive ray shooting simulation to find out the importance of multiple clumps of matter in producing strong lensing, by a hugely faster Monte Carlo technique. We showed that the probabilities that a single plane on the line-of-sight is supercritical can be approximately obtained by a simple analytical calculation. We further demonstrated that secondary matter clumps are unimportant at source redshifts $\sim 1$ but may contribute up to about $15\%$ of the single-plane cases at higher redshifts. The chance accumulation of significant amounts of secondary matter along the line-of-sight will therefore result in systematic inaccuracies in the determination of cluster masses, typically overestimating them by about $5\%$ with in addition a $\sim 5-6\%$ standard deviation. We took a simplified model of the matter distribution inside the beam and were able to cast the problem of the determination of angular diameter distances into a multiple lensing problem. We could then use our analytic model to estimate the dispersions in angular diameter distances due to inhomogeneities in the matter field. As a cautionary note, we would like to add that the for lensing involving small angular scales, the role of baryons become important and the predictions from our model in this regime, should therefore be looked at as some kind of a lower limit. An important future improvement to our model would therefore be the inclusion of the effect of baryons. It should be noted that one important advantage of our method over the halo model approach is the ability to compute strong lensing by multiple lens planes, which is rather difficult in the halo approach. Therefore, our approach provides a complementary analytic tool to the halo model based calculations, in the sense that our model will be useful to examine the effect of secondary matter when combined with the halo model approach. 
\par
There are several ways in which the power of this model could be harnessed. As noted briefly at the end of Section~\ref{sec:model}, this model may pave the way for a more accurate description of the {line-of-sight-integrated convergence} in the small angle regime where the lognormal approximation is sometimes inadequate. This issue is now in progress and will be presented elsewhere. It will also be of interest to study whether a similar model describes the 3-D PDF of dark matter density at very small scales. As the model is fixed entirely by the cosmology, one could test how efficiently it can be used to probe cosmological parameters. One of the several possible studies along these lines could be to determine how the predictions are altered by a different model of dark energy like the quintessence \citep{caldwell_etal_1998}.  With the non-Gaussianities in the lensing magnification well described by this model, one can use the scatter in Type IA supernovae to better constrain parameters like $\sigma_8$ \citep{dodelson_vallinotto_2005, cooray_etal_2005} or use multiple lens systems to better estimate the Hubble parameter $H_0$ \citep{fassnacht_etal_2005}. \par
The statistics of gravitational lensing encodes the statistical features of matter distribution and hence probes cosmology at both linear and nonlinear scales. We expect that the model proposed in this paper will be able to play an important role in  learning more about cosmology in a fashion complementary to that derivable from the analysis of large numerical simulations. The latter, of course, permit more detailed questions to be raised and answered than the analytical approach presented here. But the present approach is of sufficient accuracy and so much more computationally efficient than than based on direct simulation that we can now efficiently compare a suite of cosmological models. As an example of this, we could ask for the differences in the gravitational lensing predicted for models which give essentially identical, i.e. degenerate, cosmic microwave background signals \citep[see, for e.g.][]{huey_etal_1999,maor_etal_2001,maor_etal_2002}.

\acknowledgments{ We sincerely thank Paul Bode for providing the data from the TPM simulations. SD specifically acknowledges his help in explaining various aspects of the simulation and the data. We thank Joe Hennawi for several careful readings of the manuscript and useful comments. SD would also like to thank Masamune Oguri for useful discussions and comments. Finally, we would like to thank the referee for his thoughtful suggestions.}

\end{document}